\documentclass[a4paper,11pt]{article}
\usepackage{jcappub} 
\usepackage{lineno}
\usepackage{graphicx}
\usepackage[normalem]{ulem}

\usepackage[ngerman,english]{babel}
\newif\ifAMStwofonts
\AMStwofontstrue

\def\gsim{~\rlap{$>$}{\lower 1.0ex\hbox{$\sim$}}}

\def\simpropto{\lower.2ex\hbox{$\; \buildrel \propto \over \sim \;$}}
\def\ltsim{\lower.5ex\hbox{$\; \buildrel < \over \sim \;$}}
\def\gtsim{\lower.5ex\hbox{$\; \buildrel > \over \sim \;$}}
\def\ltsim{\lower.5ex\hbox{$\; \buildrel < \over \sim \;$}}
\def\gtsim{\lower.5ex\hbox{$\; \buildrel > \over \sim \;$}}

\def\ln{{\rm ln}}

\def\pmb#1{\setbox0=\hbox{#1}%
\kern-.025em\copy0\kern-\wd0
\kern.05em\copy0\kern-\wd0
\kern-.025em\raise.0433em\box0}

\def\simlt{\lower.5ex\hbox{$\; \buildrel < \over \sim \;$}}
\def\simgt{\lower.5ex\hbox{$\; \buildrel > \over \sim \;$}}

\newcommand{\beq}{\begin{equation}}

\newcommand{\eeq}{\end{equation}}
\def\beqa{\begin{eqnarray}}
\def\eeqa{\end{eqnarray}}
\def\fixit#1{}

\title{\boldmath Probing Cosmological Principle using the spectral index of quasar flux distribution}

\author[a]{Mohit Panwar}
\author[b]{Pankaj Jain}
\affiliation[a]{Department of Physics, Indian Institute of Technology, Kanpur 208016, India}
\affiliation[b]{Department of Space, Planetary \& Astronomical Sciences \& Engineering (SPASE), Indian Institute of Technology, Kanpur 208016, India}

\emailAdd{panwarmohit706@gmail.com}
\emailAdd{pkjain@iitk.ac.in}

\abstract{We study the dipole signal in the spectral index $(x)$ of the differential number counts using quasars in the CatWISE2020 catalog of infrared sources. The index is extracted by using the log-likelihood method. We obtain the value $x=1.579 \pm 0.001$ for a quasar sample of 1355352 sources. We extract the dipole signal in this parameter by employing $\chi^{2}$ minimization, assuming a sky model of $x$ up to the quadrupole term. We find that the dipole amplitude $|D|$ is $0.005 \pm 0.002$ and dipole direction $(l, b)$ in Galactic coordinate system equal to $(201.50^{\circ} \pm 27.87^{\circ}, -29.37^{\circ} \pm 19.86^{\circ})$. The direction of dipole anisotropy is found to be very close to the hemispherical power asymmetry $(l,b)=(221^\circ,-27^{\circ})$ in the Cosmic Microwave Background (CMB). We also obtain a signal of quadrupole anisotropy which is correlated with the ecliptic poles and can be attributed to ecliptic bias.}


\begin{document}
\maketitle
\flushbottom

\section{Introduction}
\label{sec:intro}
The $\Lambda$CDM model is based on the Cosmological Principle which states that the universe is statistically homogeneous and isotropic on large distance scales. Observationally, this principle is supported by the discovery of highly isotropic thermal cosmic microwave background(CMB) radiation at a temperature of $\sim 3$K \citep{1965ApJ...142..419P}. The largest deviation from the isotropy in the temperature field of the CMB radiation over the sky is due to the dipole which has an amplitude of $\sim 3$mK. This has been established firmly in the first year of data from Cosmic Background Explorer (COBE) \citep{1993ApJ...419....1K}, Wilkinson Microwave Anisotropy Probe (WMAP) \citep{Bennett_2003} and data from the Planck mission 
 \citep{Planck_Mission,Planck_Dipole}. This dipole temperature anisotropy is commonly believed to arise due to the solar system's peculiar motion with respect to the preferred frame, also known as the cosmic frame of rest, in which the cosmological principle holds. The velocity of the solar system relative to the cosmic frame of rest has been found to be $369.82$ kms$^{-1}$ $\pm$ $0.011$ kms$^{-1}$ towards the direction $(l,b)=(264.021^{\circ} \pm 0.011^{\circ},\: 48.253^{\circ}\pm 0.005^{\circ})$ \citep{Planck_2018_results}. 

In the standard Big Bang model, we expect that the distant sources (galaxies/quasars), which act as tracers of the large-scale structure of the Universe, would be distributed isotropically in the cosmic frame of rest.
Due to our motion with respect to this frame, we expect a dipole anisotropy in the number count of such sources due to the Doppler and aberration effects. The resulting dipole anisotropy is known as the kinematic dipole. It is reasonable to assume that the source spectrum follows, $S(\nu)\propto \nu^{-\alpha}$. Furthermore, the cumulative number count or the integral number count above the flux density $S$ of any class of extragalactic sources generally follows the power law,
\begin{equation}\label{eq:power_law}
    N(>S) \propto S^{-x}
\end{equation}
 where $x$ is the spectral index. Under these assumptions, the kinematic dipole \citep{Baldwin_Ellis_1984} is given by
\begin{equation}\label{Baldwin_Ellis}
    \mathcal{\Vec{D}}=[2+x(1+\alpha)]\Vec{v}/c
\end{equation}

The Dipole in the large-scale structures (LSS) has been measured widely in radio as well as infrared sky survey data \citep{1998MNRAS.297..545B,Blake_2002,Singal_2011,10.1111/j.1365-2966.2012.22032.x,Rubart,TIWARI20151,10.1093/mnras/stx1631,Bengaly_2018,Secrest_2021}. A dipole with amplitude much higher than the prediction due to the CMB dipole was detected in Ref. \cite{1998MNRAS.297..545B} using the combined catalog Green Bank survey (87GB) and Parkes–MIT–NRAO survey (PMN). They concluded that this was due to the 2 percent flux mismatch between the two catalogs. After applying a peculiar flux-matching procedure, they still found a higher dipole than expected. Ref. \cite{Blake_2002} extracted the dipole using the \textbf{N}RAO \textbf{V}LA \textbf{S}KY \textbf{S}urvey(NVSS) radio catalog \citep{NVSS_catalog}, and found a $1.5\sigma$ deviation from the CMB predicted dipole. They concluded that the dipole is consistent with the kinematic interpretation. In contrast, using the same catalog, \cite{Singal_2011} claimed that the dipole amplitude in number counts is nearly four times larger than the expected kinematic dipole, while the direction was found to be same as that of the CMB dipole. This was further studied by \cite{10.1111/j.1365-2966.2012.22032.x} who reported, after taking care of several systematics, that the dipole was quite different from the expected kinematic dipole. Ref. \cite{Rubart} also extracted the dipole using NVSS and WENSS radio catalogs with the help of the simple linear estimator originally proposed by \cite{Crawford_2009}. The dipole signals in these two catalogs were found to be consistent with each other, with the dipole amplitude in NVSS approximately four times larger than the expected kinematic amplitude. 

It was pointed out by \cite{Rubart} that the integral number count from the NVSS radio catalog does not follow the simple power law. This was further studied in \cite{TIWARI20151} who analysed the NVSS data with improved power law for the differential number count i.e. number of sources per unit solid angle per unit flux. This paper also finds the dipole direction matches the CMB dipole direction, but the extracted velocity exceeds roughly three times the CMB velocity. Ref. \cite{10.1093/mnras/stx1631} also obtained the velocity of solar system with respect to cosmic rest frame using the linear estimator of \cite{Crawford_2009}, and found it to be roughly four times larger than the expected value. Ref. \cite{Bengaly_2018} studied the dipole in the TGSS data and obtained the dipole amplitude approximately five times larger than the expectation based on the CMB measurement. Such a large dipole may be attributed to large systematics in the TGSS data. Indeed large flux calibration systematics present in TGSS data make this catalog unsuitable for low multipole analysis \citep{Tiwari_2019}. 

 The dipole in large scale structure has also been studied at infrared wavelengths using the quasar sources \cite{Secrest_2021}  in the CatWISE2020 catalog \citep{Marocco_2021}. They found that the dipole amplitude is approximately two times larger than that expected from CMB dipole. The deviation from kinematic dipole has a significance of approximately $4.9\sigma$ and hence presents a major challenge to the $\Lambda$CDM model. This signal has been confirmed by other authors \cite{2023MNRAS.525..231D,Kothari:2022bjr}.  However, the theoretical prediction of the dipole using Eq. \ref{Baldwin_Ellis} does not consider the effect of evolution of the luminosity function. This should be included in order to assess the significance of the excess dipole amplitude \citep{Guandalin_2023}. Furthermore, the signal appears to be consistent with $\Lambda$CDM \cite{2023MNRAS.tmp.3599M} in the Quaia data sample \cite{2023arXiv230617749S}. 

Many other observations indicate a potential departure from $\Lambda$CDM model. These include a dipole in the radio polarization offset angles \citep{Jain1998}, alignment of quasar polarization at optical frequencies at Gpc distance scales \citep{Hutsemekers_1998,Hutsemekers_Lamy_2001}, alignment of radio polarizations  \citep{Tiwari_Jain_2013,Pelgrims_Hutsemekers_2015,Tiwari_Pankaj_2016}, alignment of radio axes \citep{Taylor,Panwar} and anisotropy in the Hubble constant \cite{PhysRevD.105.103510}. Ref. \cite{Park_Chen-Gyung} complied the sample of Luminous Red Galaxies (LRG) from the Sloan Digital Sky Survey to test the homogeneity and found the Cosmological Principle does not hold up to the length scale of $300h^{-1}$Mpc. Furthermore, large scale bulk flow observations are in significant tension with 
$\Lambda$CDM \cite{Kashlinsky_2008,10.1093/mnras/stad1984}. There also exist several anomalies in the CMB data, such as, the hemispherical power asymmetry \citep{Eriksen_2004_hemispherical_power_asymmetry}, alignment of quadrupole and octopole harmonics \citep{Quadrupole_and_octopole_alignment,Octo_Quadrupole_align} and dipole modulation in the CMB polarization \citep{Ghosh_2016}. The hemispherical anisotropy is found dominantly for low multipoles and can lead to significant directional dependence of cosmological parameters \cite{10.1093/mnras/stab1193,PhysRevD.105.083508}. At higher $l$, i.e. $l\ge 32$, a significant signal of cosmological parameter variation is seen in three patches \cite{10.1093/mnras/stab1193}.  Interestingly the quadrupole, octopole and dipole modulation in CMB polarization all point close to the direction of the CMB dipole \citep{Ralston2004}. 
In contrast, the CMB hemispherical power asymmetry peaks in the direction $(l,b)=(221^\circ,-27^{\circ})$ \cite{Eriksen_2004_hemispherical_power_asymmetry}. Taken collectively the CMB anomalies are very strongly significant \cite{Jones:2023ncn}.
Many of these challenges for the $\Lambda$CDM model are nicely listed in \citep{Challenges_for_Lambda-CDM_An_update,Kumar}. However, a search for a signal of dipole modulation or equivalently hemispherical anisotropy in large scale structure leads to a null result \cite{Cobos_Frenandez}.

The observations described above suggest a potential departure from the cosmological principle. In particular, it may not be possible to interpret the dipole in source number counts in terms of motion relative to the cosmic frame of rest and hence it may 
have a different, unknown, origin. In view of this it is possible that several other observables which are not affected by our motion may also acquire a dipole. Ref. \cite{Ghosh_2017} studied dipole anisotropy in power index $(x)$ in NVSS data and concluded that it was consistent with isotropy. In \cite{Kothari:2022bjr} dipole anisotropy in the parameter $\alpha$, which parameterises the source spectrum ($S\propto \nu^{-\alpha}$), and in the mean flux per source was investigated using the data from CatWISE2020 catalog \citep{Marocco_2021}. The authors reported a strongly correlated dipole signal in two parameters, i.e., the dipole in the mean flux points towards the galactic center, while the dipole in $\alpha$ points roughly opposite to the galactic center. A simple galactic extinction model explains the correlation between these dipoles. Hence \cite{Kothari:2022bjr} conclude that there is no indication of a cosmological signal in the dipoles of these two parameters.

In the present paper we study the dipole in the spectral index $(x)$, defined in Eq. \ref{eq:power_law}, which parametrizes the flux distribution of sources, using the CatWISE2020 catalog \citep{Marocco_2021}.  
This observable is not affected by our motion with respect to the cosmic frame of rest and hence a significant dipole signal in this observable directly implies cosmology beyond the $\Lambda$CDM model. Test of anisotropy in this observable in the NVSS data sample leads to a null result \citep{Ghosh_2017}, consistent with isotropy. We will also determine whether the power law provides a good fit to the flux distribution.

The paper is organized in the following manner. In section \ref{sec:DATA_Discription}, we describe the Data used for this study and the various cuts and masks to remove the known systematic effects which may bias the results. The section \ref{theory} describes the log-likelihood procedure for estimating the spectral index $(x)$. In section \ref{data_analysis}, we discuss some issues that arise due to pixelization and the extraction of dipole signal using $\chi^{2}-$minimization. Results are presented in the section \ref{results}. We conclude in section \ref{conclusions}.

\section{Data}
\label{sec:DATA_Discription}
The data consists of the full sky scan by \textbf{W}ide-field \textbf{I}nfrared \textbf{S}urvey \textbf{E}xplorer (\textbf{WISE}) \citep{Wright_2010} in four infrared bands W1, W2 W3 and W4 centred at $3.4$, $4.6$, $12$, and $22 \mu m$ respectively and is available in the form of the CatWISE2020 catalog \citep{Marocco_2021,2013wise.rept....1C,Mainzer_2011,Mainzer_2014}.  The catalog lists $1,890,715,640$ objects with signal-to-noise ratio $S/N \geq 5$ and has $90 \%$ completeness depth at $W1=17.7$ mag and $W2=17.5$ mag. We follow \cite{Secrest_2021} and select quasars from the CatWISE2020 catalog using a simple mid-infrared colour criterion, $W1-W2 \geq 0.8$ \citep{Stern_2012}, which uses the fact that a mid-infrared selection is able to identify both unobscured (type 1) and obscured (type 2) AGNs. The resulting sample of sources has been corrected for Galactic reddening using a dust map \citep{Planck_all_sky_thermal_dust_emission_model} and extinction coefficients given in \cite{Wang_2019}.

In \cite{Secrest_2021}, the authors point out a total of 291 regions of poor-quality photometry measurement over the sky. These regions include the planetary nebulae in our Galaxy and neighbouring galaxies. These areas have been masked. For neighbouring galaxies like Andromeda and Magellanic Clouds, $6$ times the size of the $20$ mag arcsec$^{-2}$ isophotal radii from the 2MASS Large Galaxy Atlas regions have been masked. To remove the image artifacts near the bright sources in WISE data, a circular
mask with 2MASS K-band-dependent radii $\log_{10}(r\deg^{-1})=- 0.134\:K - 0.471$, is used.
The WISE scanning pattern which goes in great circles  from the north ecliptic pole to the south ecliptic pole, with center located at the sun, makes highly redundant coverage at the ecliptic poles. Any systematics arising from this apparent scanning strategy is called ecliptic bias. This bias may be responsible for the quadrupole moment seen in number counts \citep{Kothari:2022bjr}.
 A magnitude cut of $9 > W1 > 16.4$ is applied to remove faint and very bright sources. We also need to apply a Galactic plane cut, i.e., mask region corresponding to $|b|<30^{\circ}$, to remove galactic contamination \citep{Secrest_2021,Kothari:2022bjr}. The sources with low mean coverage depth W1Cov, W2Cov $<80$ also have been excluded from the data. In our final data, we are left with $1355352$ quasar candidates. This quasar sample is the same as used by \cite{Secrest_2021} to extract the dipole in number count. In \cite{Secrest_2022} the authors estimated the spectral index $x=1.89$ of integral number count in same quasar candidates of CatWISE2020 data with magnitude cut $W1<16.5$ $(S>0.078$ mJy) instead of $W1<16.4$ $(S>0.085$ mJy).

\section{Theory}\label{theory}
We assume a power law distribution of the 
 cumulative number count above a flux limit $S$, as given in Eq. \ref{eq:power_law}.
The corresponding differential number counts $n(\theta, \phi, S)$, which is defined as the number of sources per unit flux per unit solid angle at flux $S$, is given by the following expression,
\begin{equation}
    n(\theta, \phi, S) = \bigg[\frac{d^2N}{d\Omega dS}\bigg ] = kS^{-1-x}
    \label{eq:ndist}
\end{equation}
Here $k$ is the normalization constant which depends on the total number of sources.

We use a maximum likelihood method, developed earlier in \cite{1970ApJ...162..405C} and \cite{Ghosh_2017}, to estimate the spectral index (x). Let $P(S|x)$ be the probability of observing a quasar with flux density $S$ in the sky region $\Omega_{s}$. It can be expressed as,
\begin{equation}
    P(S|x)=\frac{1}{N_{T}}\int_{\Omega_{s}}n(\theta, \phi, S)d\Omega = \frac{k \Omega_{s}}{N_{T}}S^{-1-x}
\end{equation}
where $N_{T}$ is given by 
\begin{equation}
    N_{T} = \int_{\Omega_{s}}\int_{S_{min}}^{S_{max}}n(\theta, \phi,S)d\Omega dS = \frac{k\Omega_{s}}{x}(S_{min}^{-x}-S_{max}^{-x})
\end{equation}
The likelihood function $L(S|x)$ is defined by $\prod\limits_{i=1}^{N_{T}}P(S_{i}|x)$. Hence, we can write the log-likelihood function as,
\begin{equation}
    \ln L = \ln \prod\limits_{i=1}^{N_{T}}P(S_{i}|x) = N_{T} \ln\frac{x}{S_{min}^{-x}-S_{max}^{-x}} + (-1-x)\sum_{i=1}^{N_{T}}\ln S_{i}
\end{equation}

We can also extend the above analysis for a differential modified power law \cite{TIWARI20151}, given by
\begin{equation}
    n(\theta, \phi, S) = \bigg[\frac{d^2N}{d\Omega dS}\bigg ] = kS^{-1-x-\beta\log S}
    \label{eq: ndist_modified}
\end{equation}
This was found to provide a much better fit to the NVSS data in comparison to a simple power law \cite{TIWARI20151}.
In this case the probability $P(S|x, \beta)$ is given by
\begin{equation}
    P(S|x, \beta) = \frac{S^{-1-x-\beta\log S}}{\int_{S_{min}}^{S_{max}}S^{-1-x-\beta\log S}dS}
\end{equation}
The log-likelihood for this modified power law is given by
\begin{equation}
    \log L = -N_{T}\log\Big[\int_{S_{min}}^{S_{max}}S^{-1-x-\beta\log S}\Big] - \sum_{i=1}^{N_{T}}(1+x+\beta\log S_{i})\log S_{i}
\end{equation}

The value of the spectral index $(\Bar{x})$ and $\Bar{\beta}$ that maximizes the log-likelihood function are best model parameters. The uncertainty in spectral index $(\Bar{x})$ and in $\Bar{\beta}$ are given by, 

\begin{equation}
    \sigma_{x} = \bigg[-\frac{d^{2}\ln L}{dx^{2}}\bigg |_{x=\Bar{x}}\bigg]^{-1/2}
\end{equation}

\begin{equation}
    \sigma_{\beta} = \bigg[-\frac{d^{2}\ln L}{d\beta^{2}}\bigg |_{\beta=\Bar{\beta}}\bigg]^{-1/2}
\end{equation}

\begin{figure}[h]
\centering 
\includegraphics[width=7.6cm]{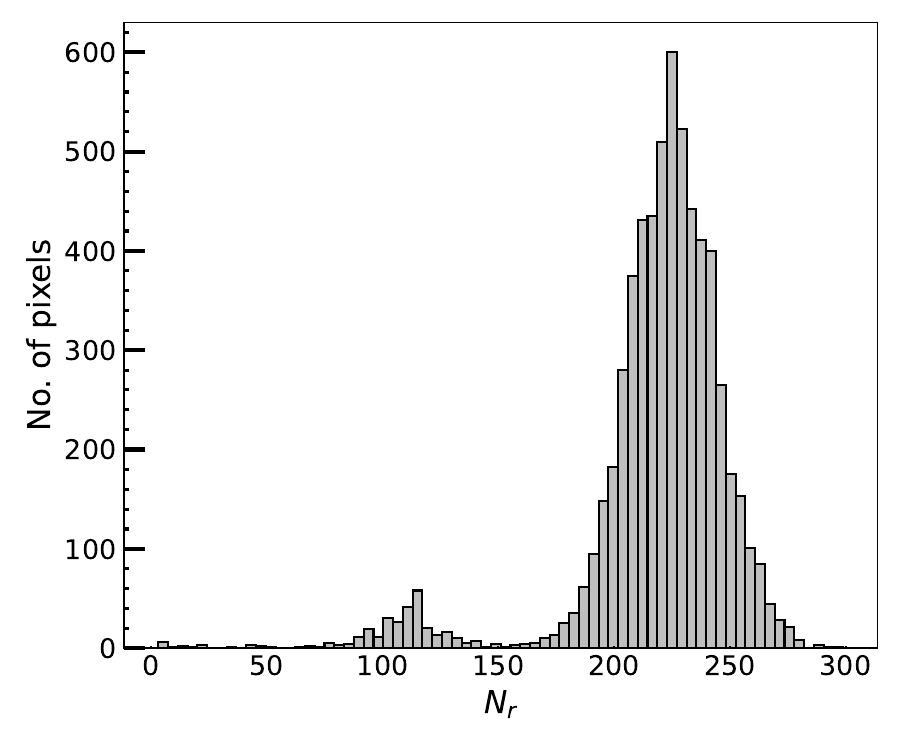}
\hfill
\includegraphics[width=7.6cm]{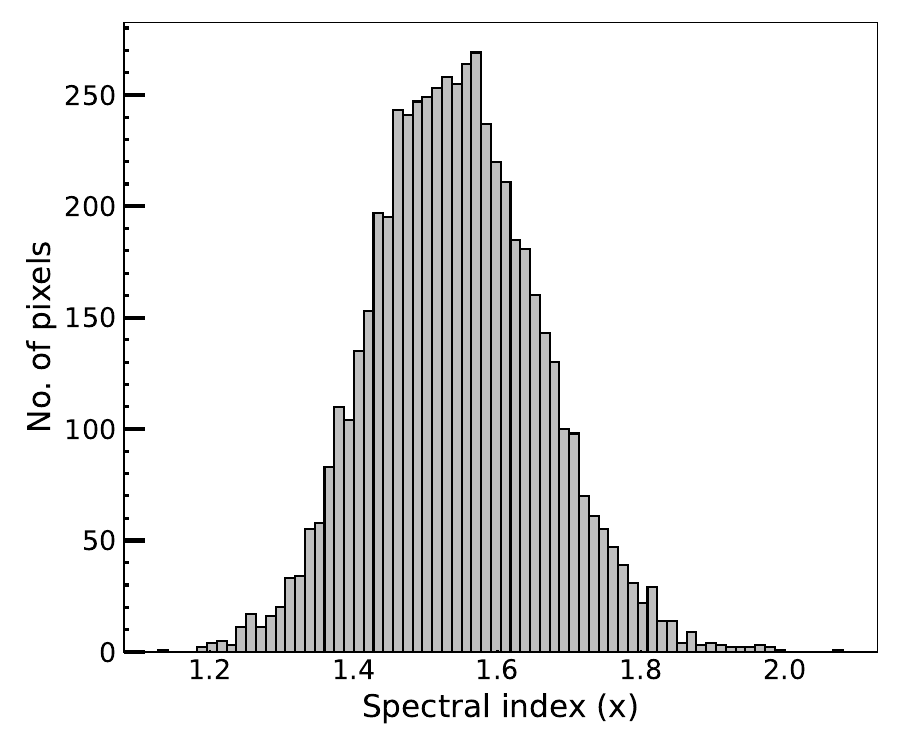}
\caption{\label{fig:ncount_dist} The  histograms of the number count ($\textit{left}$) and the spectral index $(x)$ ($\textit{right}$) using the healpy pixelization. In the case of $x$, only those pixels are included which have number counts greater than $150$.}
\end{figure}

\section{Data Analysis} \label{data_analysis}
\begin{figure}[h]
\centering 
\includegraphics[width=13cm]{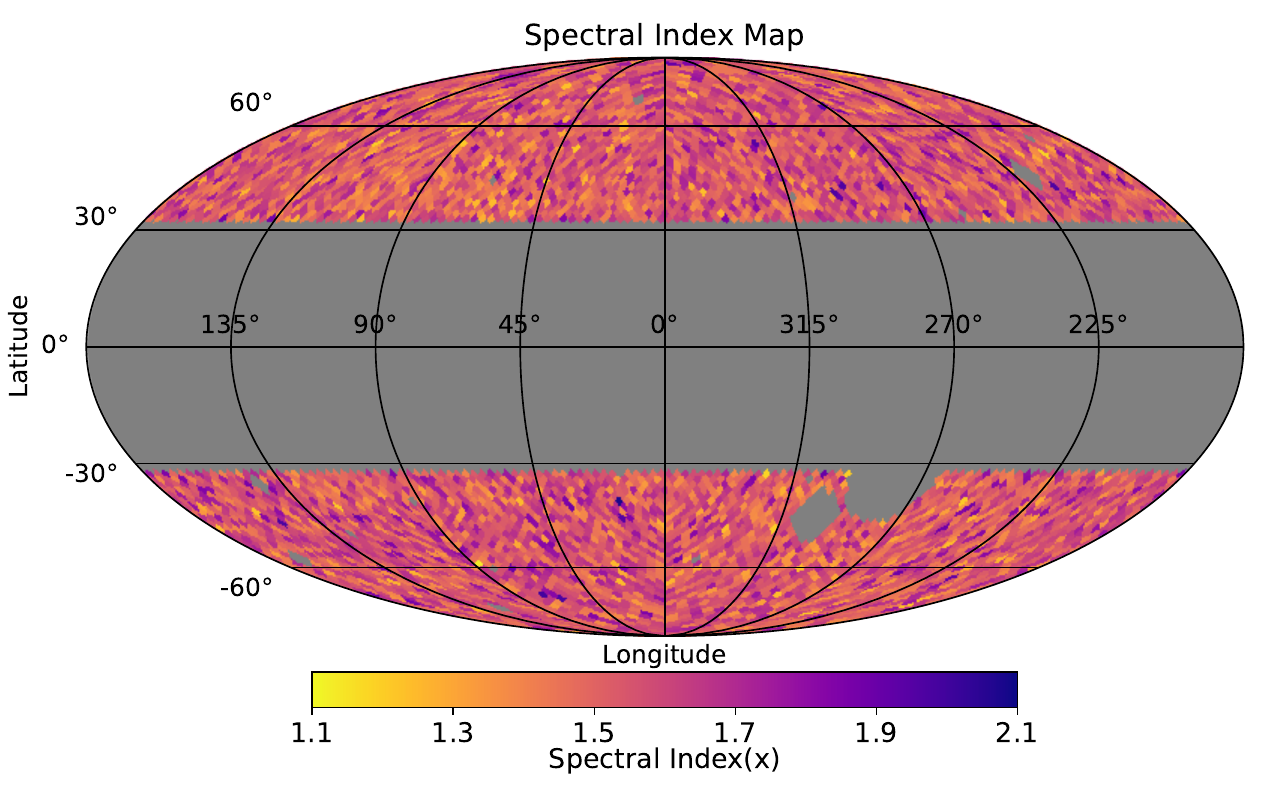}
\caption{\label{fig:x_map} Spectral index map in Galactic coordinate system}
\end{figure}

The maximum log-likelihood estimation (MLE) of spectral index (x) is implemented numerically using the \texttt{Python} based \texttt{scipy} package function \texttt{scipy.optimize.minimize()} with \texttt{Nelder-Mead} algorithm. 
To obtain the signal of dipole anisotropy in the spectral index, we divide the sky into 49152 equal-area pixels by setting Nside=64 using the healpy pixelization scheme.
The left figure of Fig. \ref{fig:ncount_dist} shows the histogram of the number counts in the pixels. From this plot, we can spot two peaks which are separated by the number count 150. It turns out that the pixels with a number count of fewer than 150 lie on the boundary of the masked regions. Some of these pixels may have very few data points and can lead to bias in the final result. Hence we have masked the pixels with a number count of less than 150 from our analysis. We have also masked all neighbouring pixels of the masked pixels. The resulting sky coverage becomes $45.65\%$. We estimated each unmasked pixel's spectral index $x$ by maximizing the log-likelihood function and extracting the one-sigma error in the spectral index. The right figure of \ref{fig:ncount_dist} shows the histogram of the estimated spectral indices of the unmasked pixels. Figure \ref{fig:x_map} shows the sky map of spectral indices in the Galactic coordinate system. 
We extract the multipole parameters using $\chi^{2}$-minimization,
\begin{equation}
    \chi^2=\sum_{p=1}^{N_{t}}\bigg[\frac{x_{F}(p)-x_{O}(p)}{\sigma_{x_{p}}}\bigg]^{2}
    \label{eq:chi2}
\end{equation}
where $x_O(p)$ is the observed value of the index $x$ in the pixel $p$ and $x_F(p)$ is the corresponding
value of the fitting function. Here the  
sum is over all unmasked pixels, and $N_{t}$ is the total number of such pixels. 
We make a multipole expansion of the sky dependence of the fitting function $x_F(\hat n)$  and keep terms up to the
quadrupole because the power associated with higher multipoles is negligible. Hence, in any pixel $p$, this function
takes the form,
\begin{equation}
    \begin{split}
    x_F(p)=M_{0}+n_{x_{p}}D_{x}+n_{y_{p}}D_{y}+n_{z_{p}}D_{z}
    + Q_{xy}n_{x_{p}}n_{y_{p}}+Q_{zy}n_{y_{p}}n_{z_{p}}+Q_{xz}n_{x_{p}}n_{z_{p}}\\
    + Q_{x^{2}-y^{2}}(n_{x_{p}}^{2}-n_{y_{p}}^{2}) + Q_{z^2}(3n_{z_{p}}^2-1)\\
    \end{split}
\end{equation}
where $M_{0}$ is Monopole, $D_{x},\:D_{x},\:D_{z}$ are the dipole parameters and $(Q_{xy},Q_{zy},Q_{xz},Q_{x^{2}-y^{2}} Q_{z^2})$ are the quadrupole parameters. Here, $(n_{x_{p}}, n_{y_{p}}, n_{z_{p}})=(\cos\theta_{p}\cos\phi_{p},\cos\theta_{p}\sin\phi_{p},\sin\theta_{p})$ are the Cartesian coordinates of the $p^{th}$ pixel.
We estimate all fitting parameters by minimizing Eq. \ref{eq:chi2}.We define the normalized dipole amplitude as,
\begin{equation}
    |D| = \frac{\sqrt{D_{x}^2+D_{y}^2+D_{z}^2}}{M_{0}}
\end{equation}

\section{Results and Discussion}\label{results}
\begin{figure}[h]
\centering 
\includegraphics[width=7.6cm]{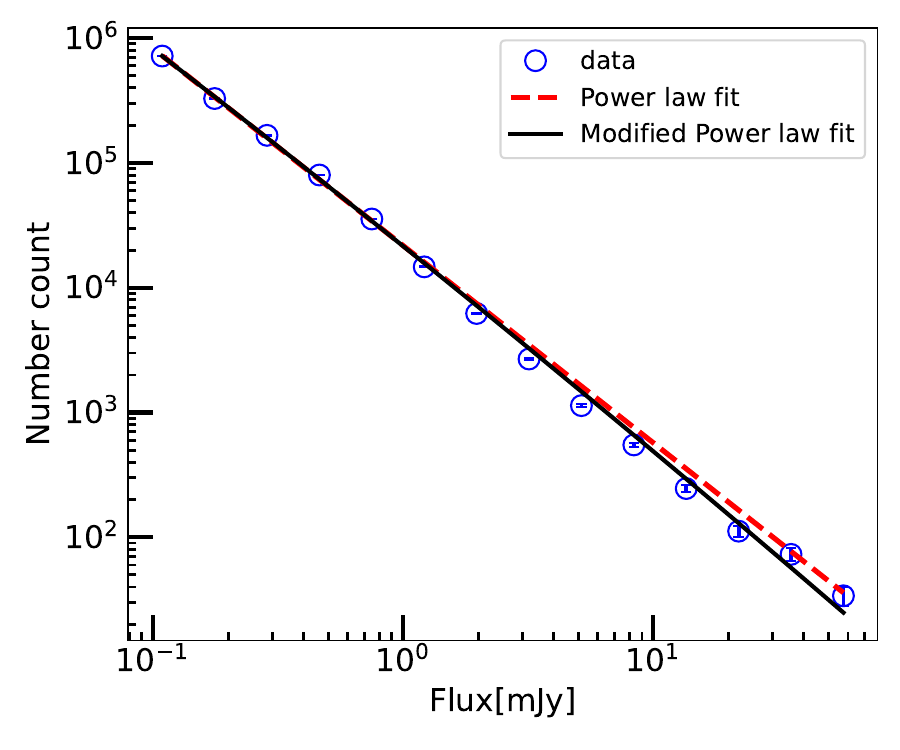}
\hfill
\includegraphics[width=7.6cm]{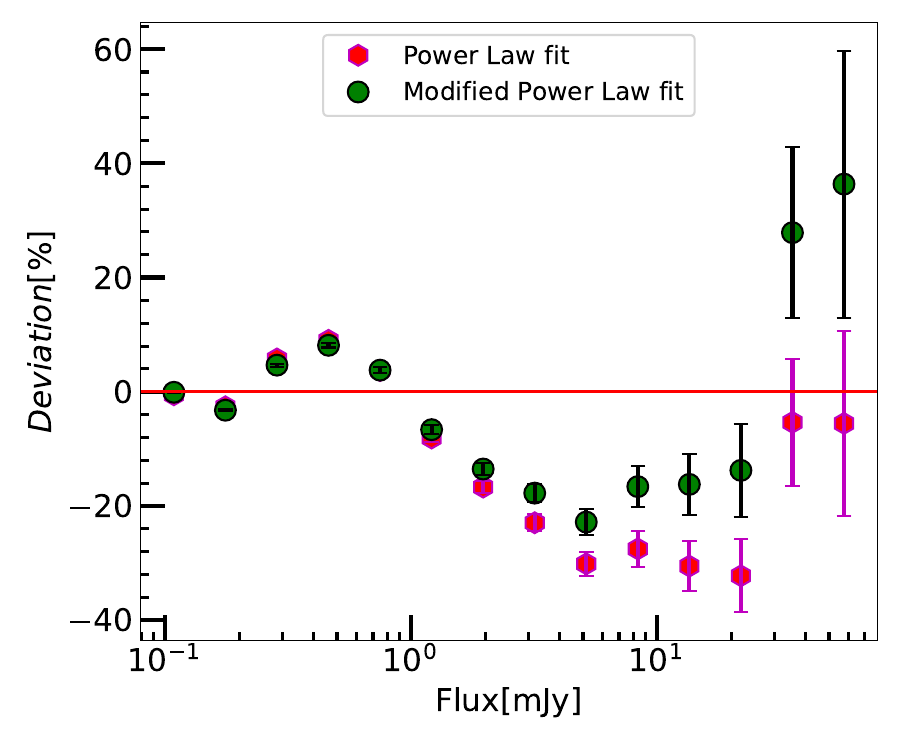}
\caption{\label{fig:powerfit} $\textit{Left:}$ A plot of the number count in equal bin width in $\log_{10}$(S/mJy) calculated from the differential power law $[n(S)\propto S^{-1-x}]$ and differential modified power law $[n(S)\propto S^{-1-x-\beta \log S}]$ model for the differential number count for our CatWISE2020 catalog data sample. The differential power law exponent has been estimated by maximizing the corresponding Log-Likelihood functions as described in section \ref{theory}. $\textit{Right}$ A plot of the deviation $\Big[\frac{Data-Model}{Model}\times 100\Big]$ of the data from the Power law and Modified power law.}
\end{figure}

Using the likelihood analysis, we find that the differential number count of our sample of $1355352$ sources follows the power law distribution given in Eq. \ref{eq:ndist} with spectral index $x=1.579 \pm 0.001$. In the left of the Figure \ref{fig:powerfit}, we compare the best-fit power law distribution with data. We find that data deviates from the power law which indicates that the whole population of our sample cannot be described with the power law. The right panel of Fig. \ref{fig:powerfit} illustrates that this deviation varies systematically with flux and is significant. From the figure we see that $x$ is expected to be large for low flux sources, followed by a decrease and finally an increase in $x$.

We also try to model the population with generalized or modified power law as given by Eq. \ref{eq: ndist_modified}. The best-fit parameters in this case are $x=1.613\pm 0.001$ and $\beta = 0.014\pm 0.001$. From Figure \ref{fig:powerfit}, we see that the modified power law exhibits a trend similar to that of a power law. However, as the flux density exceeds 1.0 mJy, the deviation of the data from the model decreases, although significant systematic variations still persists. If the extragalactic source population does not follow a power law then the theoretical predictions for the dipole based on Eq. \ref{Baldwin_Ellis} would depend on the lower cutoff on flux and may not be very reliable. In the present case this problem persists even with the modified power law \citep{TIWARI20151}. The effect of variation in $x$ on the number counts dipole requires a detailed study which we do not pursue in the present paper.

For the entire data set, we find a signal of dipole in the spectral index $x$, with dipole amplitude $|D|$ equal to $0.005 \pm 0.002$ and dipole direction $(l, b)$ in Galactic coordinate system $(201.50^{\circ}\pm 27.87^\circ, -29.37^{\circ}\pm 19.86^\circ)$. The signal has a significance level of 2.5 sigmas. Surprisingly, we find that the dipole anisotropy axis is close to the direction of hemispherical power asymmetry $(l,b)=(221^\circ,-27^{\circ})$ in CMB \cite{Eriksen_2004_hemispherical_power_asymmetry}. The two agree with one another within errors.  Essentially we obtain a larger spectral index in this direction which indicates a relatively larger number density of sources with lower flux.  We also extract the dipole in $x$ using the \texttt{Healpy fit\_dipole} method. In this case, dipole amplitude $|D|$ equals to $0.006$ and dipole points in direction $(l,b)$,  $(227.59^\circ, -34.43^\circ)$ in the Galactic coordinate system. The  \texttt{Healpy fit\_dipole} method does not consider the errors in the $x$ values for each pixel, but within the error bars, results from both methods agree with each other. We also expand $x_{p}(theoretical)$ up to the dipole term only and obtain the dipole amplitude $|D|=0.006\pm 0.002$ and dipole direction $(l,b)=(224.19^{\circ}\pm 25.0^\circ, -30.76^{\circ}\pm 18.29^\circ)$. This also agrees well with the values obtained by other methods.

\begin{figure}[h]
\centering 
\includegraphics[width=13cm]{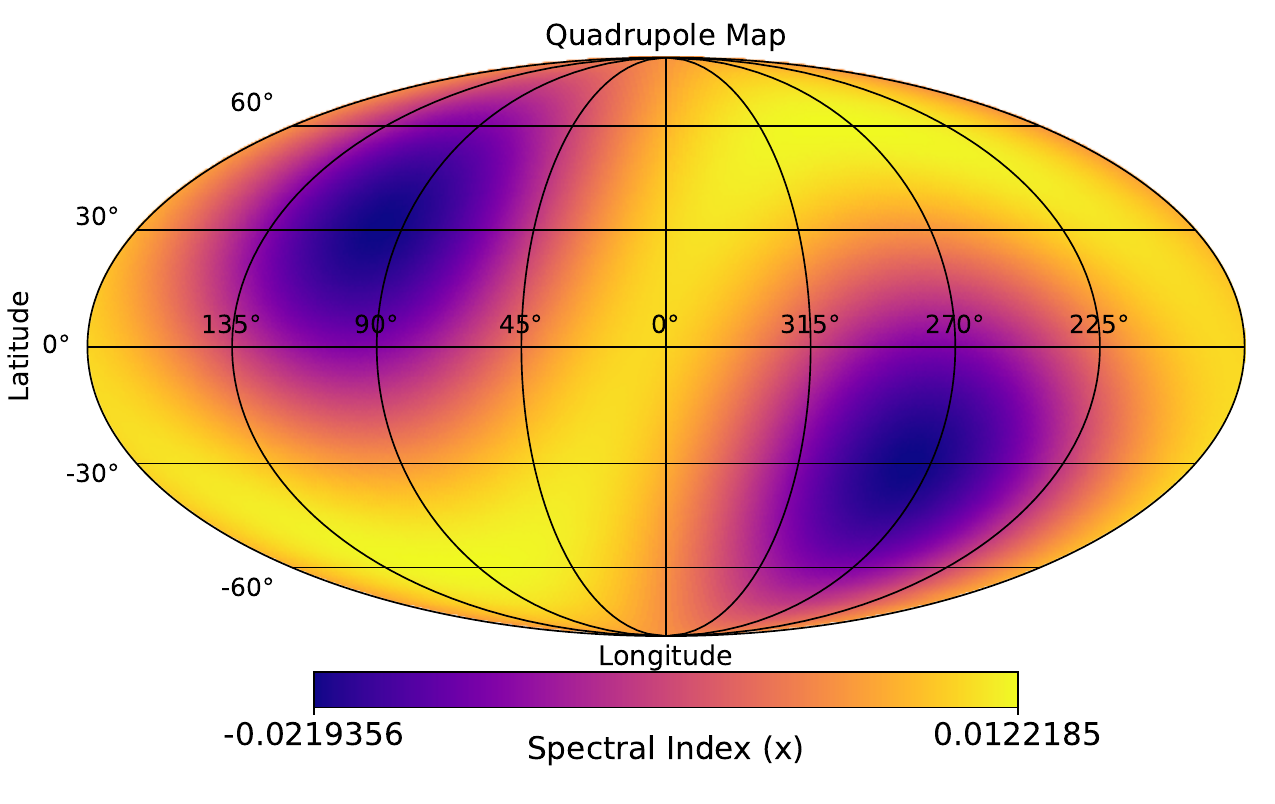}
\caption{\label{fig:Q_map} Quadrupole map of Spectral index (x) in Galactic coordinate system}
\end{figure}

We also obtain a strong signal of quadrupole moment in the spectral index $(x)$. The angular power is $C_l$ defined as,
\begin{equation}
C_{l}=\frac{1}{2l+1}\sum^{m=l}_{m=-l}|a_{lm}|^{2}    
\end{equation}
where $a_{lm}$ are the coefficients of the expansion of $x(\theta,\phi)$ in terms of spherical harmonics. The power 
 associated with Quadrupole moment $(l=2)$ is found to be $(5.6 \pm 1.3)\times 10^{-4}$ which is $\sim 6.2$ times larger than the angular power of the dipole moment $(l=1)$. The preferred or symmetry axis points towards $(l, b)=(97.03^{\circ}, 30.35^{\circ})$. The resulting quadrupole map in Galactic coordinate system is shown in Fig. \ref{fig:Q_map}. We find that the symmetry axis is aligned close to the ecliptic poles with the spectral index taking smaller values towards the poles. A similar quadrupole was also seen in the number counts \citep{Kothari:2022bjr,Secrest_2022}. In that case it was seen that the number density of the sources is less along the ecliptic poles. In the present case we obtain a smaller spectral index in these directions which is consistent with a relatively smaller density of sources corresponding to lower flux. 
The quadrupole probably arises due to some bias since it is correlated with the ecliptic poles. However,
it is not clear exactly how such a bias arises in the data sample. The bias may be attributed to higher observation time towards the
poles. In that case we would have expected comparatively more sources with low flux and hence a higher spectral index. However,
we find exactly the opposite in data. Both the number of sources and the spectral index are small along the poles. An explanation for the bias has been provided in \cite{Secrest_2022}, however, this needs further testing with more observations.

We next study the anisotropy in $x$ in the data corresponding to low flux values. For an upper limit of 0.195 mJy on $S$ we find the dipole amplitude to be $0.0072\pm 0.0058$ with best fit direction $l=146\pm 57$ and $b=-27\pm4 0$. Hence we do not obtain a significant signal. The number of sources with this cut are 974991 which is comparable to the number in the full data set. The quadrupole, however, is present in this data set with enhanced amplitude. Hence, the bias leading to the quadrupole seems to be dominant at low flux values. However, it is very interesting that the signal of dipole is absent at low flux. It is present prominently for sources with higher flux and, hence, may be less affected by systematics in data. 

\section{Conclusions}
\label{conclusions}
In this paper, we have used the \textbf{W}ide-field \textbf{I}nfrared \textbf{S}urvey \textbf{E}xplorer (\textbf{WISE}) \citep{Wright_2010} data to test for anisotropy in the power law index $x$ which parameterizes the flux distribution of sources, as given in Eq. \ref{eq:power_law}. The index is determined using likelihood analysis \citep{Crawford_2009, Ghosh_2017}. The power law model captures the general trend seen in data but also shows a significant deviation in the form of an oscillation at higher flux values (Fig. \ref{fig:powerfit}). The origin of this behaviour is so far unknown. A generalized power law distribution is also unable to provide a good fit to data in contrast to the NVSS data which prefers such a distribution \citep{TIWARI20151}. 

We test for anisotropy in the index $x$ by making a multipole expansion of this observable up to $l=2$. We find a signal of anisotropy with amplitude equal to $0.005 \pm 0.002$. Interestingly the direction of the dipole,  $(l,b)=(201.50^{\circ}\pm 27.87^\circ, -29.37^{\circ}\pm 19.86^\circ)$ is found to be in good agreement with the direction of hemispherical anisotropy or equivalently dipole modulation in CMB data \citep{Eriksen_2004_hemispherical_power_asymmetry}. We point out that the observable $x$ does not get affected by our local motion and hence any signal of anisotropy in this parameter can only be attributed to a fundamental anisotropy in the Universe. Furthermore, the signal is found to be absent for low flux sources and arises primarily from sources of higher flux which may be less affected by observational bias. The fact that the signal is in the same direction as the hemispherical anisotropy in CMB provides some support that it may be of physical origin. However, this needs further testing to rule out the possibility of a statistical fluctuation.

We also find a signal of quadrupole anisotropy with the quadrupole axis aligned with the ecliptic poles. A similar quadrupole is also seen in the case of number counts and can be attributed to the scanning pattern of the telescope. In the present case the quadrupole anisotropy is found to be dominant for sources of low flux. This is in contrast to dipole anisotropy in $x$, which is absent for the low flux sources, and hence may be less affected by observational bias.

Physically, the CMB hemispherical anisotropy or dipole modulation in CMB indicates an anisotropic distribution in the CMB power which implies that the power spectrum at the time of decoupling depends on the direction of observation. Hence this can potentially lead to an anisotropic power also in the large scale structures. Search for such a signal in NVSS data leads to a null result \citep{Cobos_Frenandez}. Our analysis suggests that there is indeed some dependence of the distribution of sources as a function of direction in the CatWise2020 catalog. Hence, it will be very interesting to search for the signal of dipole modulation of power in this catalog. It will also be interesting to theoretically evolve the anisotropic power spectrum to predict the matter power spectrum and flux distribution of sources at redshift values ranging from 1 to 10. Such a study can provide the relationship between the observed CMB power spectrum and matter distribution and hence would be very useful in making a definitive test of anisotropy.


\acknowledgments
We acknowledge the use of the \texttt{Python} packages \texttt{matplotlib}\citep{matplotlib_package} \texttt{scipy}\citep{scipy_package} and \texttt{healpy}\citep{healpy_package} for this analysis. We thank Ruth Durrer for a useful communication and Prabhakar Tiwari for useful discussions.


\bibliographystyle{JHEP}
\bibliography{biblio}

\end{document}